\documentstyle[12pt,epsf]{article}
\begin{document}
\renewcommand{\thetable}{\Roman{table}}
\def \beq{\begin{equation}}
\def \bk{\bar K^0}
\def \eeq{\end{equation}}
\def \k{K^0}
\def \m{{\cal M}}
\rightline{DOE/ER/40561-222-INT95-17-07}
\rightline{EFI-95-51}
\rightline{hep-ph/9508298}
\vspace{0.5in}
\centerline{\bf TABLE-TOP TIME-REVERSAL VIOLATION
\footnote{To be submitted to Am.~J.~Phys.}}
\vspace{0.5in}
\centerline{\it Jonathan L. Rosner}
\centerline{\it Institute for Nuclear Theory}
\centerline{\it University of Washington, Seattle, WA 98195}
\bigskip
\centerline{and}
\bigskip
\centerline{\it Enrico Fermi Institute and Department of Physics}
\centerline{\it University of Chicago, Chicago, IL 60637
\footnote{Permanent address.}}
\bigskip

\centerline{\bf ABSTRACT}
\medskip
\begin{quote}
Many electrical and mechanical systems with two normal modes are appropriate
for illustrating the quantum mechanics of neutral kaons.  The illustration
of CP- or time-reversal-violation in the neutral kaon system by mechanical or
electrical analogues is more subtle.  Some possibilities which could be
realized in a table-top demonstration are suggested.
\end{quote}
\bigskip

\centerline{\bf I.  INTRODUCTION}
\bigskip

The problem of two coupled degenerate systems is one of the first a student
encounters when learning about normal modes.  In quantum mechanics, a fine
example (see, e.g., Ref.~\cite{Baym}) is provided by the states of neutral
kaons \cite{GP}.  There exist two states $K^0$ and $\bar K^0$ which are
degenerate as long as the product of charge conjugation $C$, space inversion
$P$ (for parity), and time reversal $T$ leaves the theory invariant.
(Lorentz-invariant local quantum field theories indeed are invariant under the
combined product CPT \cite{CPT}.) In a theory invariant under CP \cite{VA},
the linear combination of $K^0$ and $\bar K^0$ with $CP = +1$ couples to two
pions, and is short-lived, while the other combination with $CP = -1$ does not,
and is long-lived. There are many ways to illustrate this two-state system in
classical physics, including vibrating membranes, coupled pendula, and coupled
oscillators.

The discovery of CP violation \cite{CCFT} indicated that the neutral kaon
system is somewhat more complex than the typical two-state problem.  Both
the short-lived and long-lived neutral kaon can decay to $\pi \pi$.  An
interesting challenge is to illustrate this behavior in a table-top setting
based on classical physics.  We shall assume a CPT-invariant theory, so that it
would suffice to illustrate either CP- or time-reversal-violation. In the
present article we suggest some possibilities based on electrical circuits,
utilizing effects in which the time-reversal violation is induced by means
of an external magnetic field.  Our purpose is to stimulate discussion of
further illustrations, with a possible eye to understanding the way in
which T-violation actually arises in Nature.

In Section II we describe three two-state systems:  the neutral kaons
(in a CP-invariant context), a coupled-pendulum analogue, and an electrical
analogue.  We then turn to the CP-violating problem for neutral kaons in
Sec.~III.  Some electrical analogues of the CP-violating problem are suggested
in Sec.~IV, while Sec.~V concludes.
\bigskip

\centerline{\bf II.  THE TWO-STATE SYSTEM IN A CP-CONSERVING THEORY}
\bigskip

\leftline{\bf A.  Neutral kaons}
\bigskip

In order to make sense of a class of ``strange'' particles produced strongly
but decaying weakly, Gell-Mann and Nishijima \cite{GN} in 1953 proposed an
additive quantum number, ``strangeness,'' conserved in the strong interactions
but not in the weak interactions.  The reaction $\pi^- p \to K^0 \Lambda$, for
example, would conserve strangeness $S$ if $S(\pi) = S(p) = 0$, $S(\Lambda) =
-1$, and $S(K^0) = +1$.  The kaon could not be its own antiparticle; there
would have to also exist a $\bar K^0$ with $S(\bar K^0) = -1$.  It could be
produced, for example, in the reaction $\pi^- p \to K^0 \bar K^0 n$.

The states $K^0$ and $\bar K^0$ would be degenerate in the absence of coupling
to final states (or to one another).  However, both states can decay to the $2
\pi$ final state in an S-wave (orbital angular momentum $\ell = 0$). Gell-Mann
and Pais \cite{GP} noted that since $C(\pi^+ \pi^-)_{\ell = 0} = +$, $C(K^0) =
\bar K^0$, and $C(\bar K^0) = K^0$, the linear combination of $K^0$ and $\bar
K^0$ which decayed to $\pi^+ \pi^-$ had to be $K_1 \equiv (K^0 + \bar
K^0)/\sqrt{2}$. Then there should be another state $K_2 \equiv (K^0 - \bar
K^0)/\sqrt{2}$ forbidden to decay to $\pi^+ \pi^-$, and thus long-lived.  (It
should be able to decay to $3 \pi$, for example.) This state was looked for
and found \cite{KL}. Its lifetime was measured to be about 600 times that of
$K_1$.

Gell-Mann and Pais assumed that C was conserved in the weak decay process.  In
a CP-conserving weak interaction theory in which C and P are individually
violated, the above argument can be recovered by replacing C with CP
\cite{KCP}.  If one chooses the phase of the particle states in such a
way that $\bk = CP \k$, the eigenstates with positive and negative CP are then,
as before,
\beq \label{eqn:k1k2}
K_1 = \frac{\k + \bk}{\sqrt{2}}~~~,~~K_2 = \frac{\k - \bk}{\sqrt{2}}~~~.
\eeq
\bigskip

\leftline{\bf B.  Mechanical analogues}
\bigskip

The $K^0 - \bar K^0$ system resembles many coupled degenerate problems in
classical physics.  For example, a drum-head in its first excited state
possesses a line of nodes. A degenerate state exists with the line of nodes
perpendicular to the first, but nothing specifies the absolute orientations of
the two lines of nodes.  However, if a fly alights off-center on the drum, it
will define the two lines of nodes.  One mode will couple to the fly, thereby
changing in frequency, and one mode will not.

A system of two coupled pendula \cite{BW} also provides a simple analogy to the
$K_1 - K_2$ system.  The requirement of CPT invariance is satisfied by taking
the two pendula to have equal natural frequencies $\omega_0$ (and, for
simplicity, equal lengths and masses).  If one couples them by a connecting
spring, the two normal modes will consist of one with frequency $\omega_1 =
\omega_0$ in which the two pendula oscillate in phase, and another with
$\omega_2 > \omega_0$ in which the two oscillate 180$^{\circ}$ out of phase,
thereby compressing and stretching the connecting spring.  If the connection
dissipates energy, the mode with frequency $\omega_2$ will eventually decay
away, leaving only the in-phase mode with frequency $\omega_1 = \omega_0$.
\bigskip

\leftline{\bf C.  Electrical analogue}
\bigskip

A simple electrical analogue of the neutral kaon system can be constructed
using two $L-C$ ``tank'' circuits, each consisting of a capacitor $C_i$ in
parallel with an inductor $L_i$ and having resonant frequency $\omega_{0i} =
(L_iC_i)^{-1/2}$ $(i = 1,2)$. Currents $I_1$ and $I_2 = - I_1$ flow through
each circuit to ground.  The two tank circuits are coupled through an impedance
$Z_c$, so that the voltages on each circuit are related to the currents by $V_2
- V_1 = Z_c I_1 = Z_c (I_1 - I_2)/2$.  The network is shown in Fig.~1.

\begin{figure}
\centerline{\epsfysize = 3.5in \epsffile{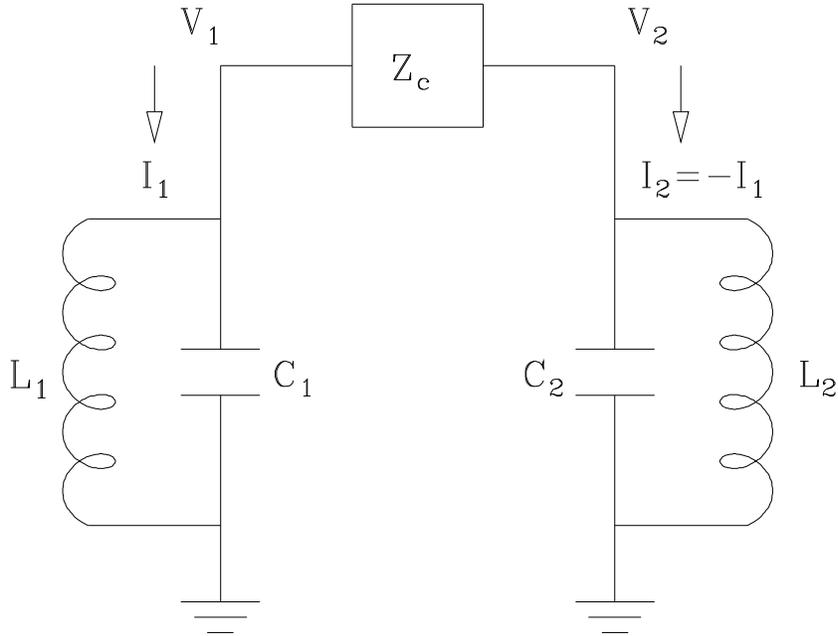}}
\caption{Coupled ``tank'' circuits illustrating the $K^0 - \bar K^0$ system.}
\end{figure}

Adding the currents flowing through each inductor and capacitor, $I_i =
I_{Li} + I_{Ci}$, and noting that $L_i \dot I_{Li} = V_i$ and $I_{Ci} = C_i
\dot V_i$, we find $\dot I_i = V_i/L_i + C_i \ddot V_i$.  Thus
\beq
\dot V_2 - \dot V_1 = \frac{Z_c}{2} \left[ \frac{V_1}{L_1} + C_1 \ddot V_1
- \frac{V_2}{L_2} - C_2 \ddot V_2 \right]~~~.
\eeq
Moreover, since $I_1 + I_2 = 0$, we can write
\beq
\frac{V_1}{L_1} + C_1 \ddot V_1 + \frac{V_2}{L_2} + C_2 \ddot V_2 = 0~~~.
\eeq
We have written the equations in a form which exhibits the symmetry
between the tank circuits 1 and 2.  Now we assume harmonic behavior:
$V_i = v_i e^{-i \omega t}$, and solve the following coupled equations in
$v_1$ and $v_2$ for characteristic values of $\omega$:
\beq \label{eqn:char1}
\left( \frac{1}{L_1} - \omega^2 C_1 - 2 i Y_c \omega \right) v_1
- \left( \frac{1}{L_2} - \omega^2 C_2 - 2 i Y_c \omega \right) v_2 = 0~~~,
\eeq
\beq \label{eqn:char2}
\left( \frac{1}{L_1} - \omega^2 C_1 \right) v_1
+ \left( \frac{1}{L_2} - \omega^2 C_2 \right) v_2 = 0~~~.
\eeq
Here we have defined the coupling admittance as the reciprocal of the coupling
impedance: $Y_c \equiv Z_c^{-1}$.

When $Y_c = 0$, if the coefficient of $v_i$ is zero but that of $v_j$ ($j \ne
i$) is nonzero, then we have a solution with $v_i\ne 0$ and $v_j = 0$.  Now,
however, let us assume the natural frequencies of the two tank circuits are the
same:  $L_1 C_1 = L_2 C_2 = \omega_0^{-2}$.  After some simplification,
the characteristic equation reduces to
\beq
(\omega_0^2 - \omega^2) \left[ 1 - \frac{\omega^2}{\omega_0^2} - i Y_c \omega
(L_1 + L_2) \right] = 0~~~.
\eeq

One solution has the natural frequency $\omega = \omega_0$ independently of the
coupling.  This is the solution with $v_1 = v_2$, where no current flows
through the coupling device.  The other solution has a frequency shift which in
general has both real and imaginary parts.  Defining $\delta \omega \equiv
\omega - \omega_0$, we find $\delta \omega/\omega \approx - 2 i Y_c (L_1 + L_2)
\omega$. A real admittance (corresponding to a resistive coupling) leads to an
imaginary frequency shift, and a damping of the oscillation.  An imaginary
admittance (corresponding to an inductive or capacitive coupling) leads to a
real frequency shift.

The solution with $v_1 = v_2$ corresponds in the neutral kaon system to the
$K_2$ state, which does not couple to two pions.  The solution with
$v_1 \ne v_2$ corresponds to the $K_1$ state, which decays to two pions and
is shifted in both mass and width from the $K_2$.
\bigskip

\centerline{\bf III.  NEUTRAL KAONS WITH CP VIOLATION}
\bigskip

The discovery \cite{CCFT} in 1964 that both the short-lived ``$K_1$'' and
long-lived ``$K_2$'' states decayed to $\pi \pi$ upset the tidy picture of a
two-state system described in Sec.~II A.  It signified that not even CP
symmetry was valid in Nature.  Henceforth the states of definite mass and
lifetime would be known as $K_S$ (for ``short'') and $K_L$ (for ``long'').
They can be parametrized approximately as
\beq \label{eqn:appx}
|S \rangle \simeq |K_1 \rangle + \epsilon |K_2 \rangle~~~,~~
|L \rangle \simeq |K_2 \rangle + \epsilon |K_1 \rangle~~~,
\eeq
where we shall use the shorthand $S,~L$ for $K_S,~K_L$. The complex parameter
$\epsilon$ encodes all we know at present about CP violation in the neutral
kaon system.  Its magnitude is $\epsilon = (2.26 \pm 0.02) \times 10^{-3}$ and
its phase is approximately 45$^{\circ}$. Notice that, in contrast to $|K_1
\rangle$ and $|K_2 \rangle$, the states $|S \rangle$ and $|L \rangle$ are not
orthogonal to one another but have a scalar product $\langle L | \rangle S
\approx 2~{\rm Re}~\epsilon$.

A convenient way to discuss the above problem is to express $K_S$ and $K_L$ as
eigenstates of a $2 \times 2$ ``mass matrix''
${\cal M}$ \cite{CL,Revs}. In the kaon
rest frame, the time evolution of basis states $\k$ and $\bk$ can be
written\cite{Sachs} as
\beq
i \frac{\partial}{\partial t}
\left[ \begin{array}{c} \k \\ \bk \end{array} \right]
= {\cal M} \left [ \begin{array}{c} \k \\ \bk \end{array} \right]~~~;
{}~~ {\cal M} = M - i \Gamma /2~~~.
\eeq
An arbitrary matrix ${\cal M}$ can be written in terms of Hermitian matrices
$M$ and $\Gamma$.  CPT invariance can be shown to imply the restriction ${\cal
M}_{11} = {\cal M}_{22}$ and hence $M_{11} = M_{22}, ~ \Gamma_{11} =
\Gamma_{22}$. We adopt this limitation here.

We denote the eigenstates of ${\cal M}$ by
\beq
|S \rangle = p | \k \rangle + q | \bk \rangle ,
\eeq
\beq
| L \rangle = p | \k \rangle - q | \bk \rangle ,
\eeq
with $|p|^2 + |q|^2 = 1$, and the corresponding eigenvalues
by $ \mu_{S,L} \equiv m_{S,L} - {i \over 2} \Gamma_{S,L}$, where
$m_{S,L}$ and $\Gamma_{S,L}$ are real.  Here we have taken the condition
${\cal M}_{11} = {\cal M}_{22}$ into account.
With $\epsilon \equiv (p-q) /( p+q)$, we can write
\beq \label{eqn:eigS}
|S \rangle = \frac{1}{\sqrt{2(1+| \epsilon |^2)}}
\left[ (1 + \epsilon ) | \k \rangle + (1 - \epsilon )| \bk \rangle \right]~~~,
\eeq
\beq \label{eqn:eigL}
| L \rangle = \frac{1}{\sqrt{2(1+| \epsilon |^2)}}
\left[ (1 + \epsilon ) | \k \rangle - (1 - \epsilon )| \bk \rangle \right]~~~.
\eeq
Expressing $\k$ and $\bk$ in terms of the CP eigenstates by means of
Eq.~(\ref{eqn:k1k2}), we recover the relation (\ref{eqn:appx}).

One can relate $\epsilon$ to the properties of the mass matrix and mass
eigenvalues. Making a phase choice, we can write
\beq
{q \over p} = \sqrt{\frac{{\cal M}_{21}}{{\cal M}_{12}}}
\eeq
and note that
\beq
\mu_S = \m_{11} + \sqrt{\m_{12} \m_{21}}~~; ~~~
\mu_L = \m_{11} - \sqrt{\m_{12} \m_{21}}~~~,
\eeq
so
\beq \label{eqn:mudiff}
\mu_S - \mu_L = 2 \sqrt{\m_{12} \m_{21}}~~~.
\eeq
Then
\beq \label{eqn:eps}
\epsilon = \frac{p-q}{p+q} =
\frac{\sqrt{\m_{12}} - \sqrt{\m_{21}}}{\sqrt{\m_{12}} + \sqrt{\m_{21}}}
\simeq
\frac{\m_{12}-\m_{21}}{4 \sqrt{\m_{12} \m_{21}}}~~~,
\eeq
where the smallness of $\epsilon$ has been used.
With the definition of ${\cal M}$ and (\ref{eqn:mudiff}) we can then write
\beq
\epsilon \simeq
\frac{{\rm Im} (\Gamma_{12}/2) + i~{\rm Im} M_{12}}{\mu_S - \mu_L} ,
\eeq
so that the CP-violation parameter $\epsilon$ arises from imaginary parts of
off-diagonal terms in the mass matrix.

The matrices $\Gamma$ and $M$ may be expressed \cite{Kabir} in terms of sums
over states connected to $\k$ and $\bk$ by the weak Hamiltonian $H_W$.
By considering specific $2 \pi,~3 \pi,~\pi l \nu$, and other final states, one
can show that $|{\rm Im} \Gamma_{12}/2| \ll |{\rm Im} M_{12}|$.
This result then implies a specific phase of $\epsilon$:
\beq
{\rm Arg}~\epsilon \approx \left\{ \begin{array}{c} 90^0 \\ 270^0 \end{array}
\right \} - ~{\rm Arg} (\mu_S - \mu_L ) ~~ {\rm for} ~~
\left \{ \begin{array}{c} {\rm Im}M_{12} > 0 \\ {\rm Im}M_{12} < 0 \end{array}
\right \}
\eeq
Given the measurements \cite{PDG} $m_S - m_L = - 0.476 ~ \Gamma_S$, $\Gamma_S -
\Gamma_L = 0.998 ~ \Gamma_S$, we have $\mu_S - \mu_L = - (0.476 + 0.499i)
\Gamma_S$, or Arg $(\mu_S - \mu_L) = (3 \pi /2) - $  arctan $(0.476/0.499) = (3
\pi/2) - 43.6^0$. Thus
$$
{\rm Arg} ~ \epsilon = (43.6 \pm 0.2)^0 ~~ ({\rm Im}~ M_{12} < 0 )~~~,
$$
\beq \label{eqn:epsph}
{\rm Arg }~ \epsilon = \pi + (43.6 \pm 0.2)^0 ~~ ({\rm Im}~ M_{12} > 0 )~~~.
\eeq

We seek in classical physics an analogue of the two-state mixing problem which
leads to a non-zero value of $\epsilon$.  Equation~(\ref{eqn:eps}) implies that
$\epsilon \ne 0$ arises from a lack of symmetry of the mass matrix $\m$.  There
may be additional manifestations of CP violation in kaon decays not reflected
in the parameter $\epsilon$ (for example, if the amplitude ratios $A(L \to \pi
\pi)/ A(S \to \pi \pi)$ differ for charged and neutral pions), but experiments
\cite{E731,NA31} do not yet conclusively demonstrate their presence.
\newpage

\centerline{\bf IV.  ELECTRICAL ANALOGUES OF T-VIOLATION}
\bigskip

\leftline{\bf A.  Mass matrix formulation}
\bigskip

In order to recast the electrical problem in a form closer to that of
Sec.~III, we rewrite Eqs.~(\ref{eqn:char1}) and (\ref{eqn:char2}) as
\beq
{\cal M} {\bf v} = \omega^2 {\bf v}~~~,
\eeq
where
\beq
{\cal M} \equiv \left[ \begin{array}{c c}
\omega_0^2 - \frac{i Y_c \omega_0}{C_1} & \frac{i Y_c \omega_0}{C_1} \\
\frac{i Y_c \omega_0}{C_2} & \omega_0^2 - \frac{i Y_c \omega_0}{C_2} \\
\end{array} \right]~~~,
\eeq
and
\beq {\bf v} \equiv \left[ \begin{array}{c} v_1 \\ v_2 \end{array}
\right]~~~.
\eeq
Here we have used the earlier assumption $L_1 C_1 = L_2 C_2 = \omega_0
^{-2}$, and have replaced $\omega$ by $\omega_0$ in the coupling term.

Note that ``CPT invariance,'' with ${\cal M}_{11} = {\cal M}_{22}$, is not
necessarily a feature of this mass matrix.  We shall assume it to be so
by taking $C_1 = C_2 \equiv C$ for simplicity.

The violation of ``CP'' or ``T'' invariance is now parametrized by allowing
the presence of small antisymmetric off-diagonal terms in ${\cal M}$:
\beq
{\cal M}_{12} \to {\cal M}_{12} - a \omega_0^2~~~,~~
{\cal M}_{21} \to {\cal M}_{21} + a \omega_0^2~~~.
\eeq
We thereby parametrize an off-diagonal coupling which is not symmetric
between $1 \to 2$ and $2 \to 1$.

The characteristic equation now becomes
\beq
(\omega_0^2 - \omega^2)^2 - 2 i \frac{Y_c \omega_0}{C} (\omega_0^2
- \omega^2) + a^2 \omega_0^4 = 0~~~.
\eeq
The solutions are modes with $\omega^2 \simeq \omega_0^2 - 2iY_c \omega_0/C$,
corresponding to $K_S$ (whose mass is affected strongly by the
``CP-conserving'' coupling $Y_c$),
and with $\omega^2 \simeq \omega_0^2 + (i/2)(a C\omega_0/Y_c)^2$, corresponding
to $K_L$ (whose mass now also receives a small contribution from the
``CP-violating'' coupling $a$).  The eigenstates
can be written in the form (\ref{eqn:eigS}) and (\ref{eqn:eigL}), if we make
the identification
\beq
\epsilon \simeq \frac{i a C \omega_0}{2 Y_c}~~~.
\eeq
Here we have assumed $a$ is sufficiently small so that $|\epsilon| \ll 1$.
The relative strength and phase of the antisymmetric and symmetric couplings
thus governs $\epsilon$, just as in the case of neutral kaons.

It is not obvious that a physical system can be constructed along the above
lines without violating ``CPT'' invariance,
since one might expect different terms to appear in ${\cal M}_{11}$ and
${\cal M}_{22}$.  The construction of an explicit ``CPT-invariant'' but
``CP-violating'' system so far has eluded us, but we now
give some possibilities.
\bigskip

\leftline{\bf B.  Propagation of VLF radio waves in the ionosphere}
\bigskip

It has been found \cite{Davies} that radio waves of very low frequencies
(in the 10 -- 20 kHz range) propagate with different phase velocities and
attenuations from west to east and east to west in the daytime ionosphere.
This behavior has been traced to interaction with electron orbits in the
Earth's magnetic field.  The variations in attenuation can amount to
differences of nearly 50\%.  This suggests that one might devise a table-top
version of non-reciprocal behavior in which the coupling unit denoted by
the box in Fig.~1 transfers energy differently from tank circuit 1 to 2
and from circuit 2 to 1.  We now suggest two ways of implementing this idea.
\bigskip

\leftline{\bf C.  Faraday rotation}
\bigskip

Let us imagine the tank circuits 1 and 2 in Fig.~1 to consist of two
oscillators with identical freqencies, coupling to each other by means of
half-wave dipole antennas whose planes of polarization are perpendicular to the
line between them.  Let each dipole be in the far (radiation) field of the
other for simplicity.  As long as the dipoles are parallel to one another, the
energy transfer between the circuits is maximal.  If the dipoles make an angle
$\phi$ with respect to one another, the power transfer will be proportional to
$\cos^2 \phi$.

Now let the dipoles be separated by a medium which causes the radio waves to
undergo Faraday rotation.  This may be achieved by letting a magnetic field be
present parallel to the line joining the two dipoles, and propagating the
radio-frequency energy through a plasma such as the earth's ionosphere. Suppose
that the medium causes a rotation of the plane of polarization by an angle
$\psi$ to the right.  This rotation will then be characteristic of radiation
traveling both from circuit 1 to circuit 2 and from 2 to 1. If antenna 2 is
oriented with respect to 1 by an angle $\phi$, then the power transfer from 1
to 2 will be proportional to $\cos^2 (\phi - \psi)$, while that from 2 to 1
will be proprtional to $\cos^2(\phi + \psi)$.  If, for example, $\phi = \psi =
\pi/4$, power will be transfered with maximum efficiency from 1 to 2, but no
power will flow from 2 to 1.

The presence of an external magnetic field with an orientation from 1 to 2
explicitly violates time-reversal invariance, so it is no surprise that one
can induce an asymmetric mixing between the two circuits.  The realization of
this system in a practical demonstration would be of some interest.  In
analogy to the mass eigenstates $|S \rangle$ and $|L \rangle$ of the neutral
kaon system, the eigenmodes would not be orthogonal to one another, and
neither eigenmode would correspond to a solution with $v_1 = v_2$.
\newpage

\leftline{\bf D.  Other possibilities}
\bigskip

Any coupling term which simulates the behavior $\m_{12} \ne \m_{21}$ in the
kaon mass matrix would be suitable for inducing an effect equivalent to
$\epsilon \ne 0$.  One could imagine utilizing, for example, the Hall effect,
in which conduction from point $1$ to point $2$ would not necessarily be the
same as that from point $2$ to point $1$ in the presence of a suitable magnetic
field. Another possibility would be to simply insert an amplifier between the
two tank circuits in Fig.~1.  In this case it is not as clear how time-reversal
invariance is violated.
\bigskip

\centerline{\bf V.  CONCLUSIONS}
\bigskip

The demonstration of CP or time-reversal violation in a classical table-top
setting would be an interesting and instructive extension of the problem of two
coupled degenerate systems, such as pendula or oscillators.  A coupling term
which is asymmetric with respect to the two systems is required.  We have
suggested some ways in which this coupling might be realized in practice.
The experience encountered in realizing such a circuit in a practical device
might well offer insights into the way in which CP- or time-reversal invariance
violation is actually realized in the neutral kaon system.
\bigskip

\centerline{\bf ACKNOWLEDGMENTS}
\bigskip

I am grateful to Bruce Winstein for first drawing my attention to table-top
models of the neutral kaon system, to Brian Fick for a useful suggestion,
and to P. G. H. Sandars for helpful comments on the manuscript. I
wish to thank the Institute for Nuclear Theory at the University of Washington
and the Aspen Center for Physics
for hospitality during this work, which was supported in part by the United
States Department of Energy under Grant No. DE FG02 90ER40560.
\bigskip

\def \ajp#1#2#3{Am. J. Phys. {\bf#1}, #2 (#3)}
\def \apny#1#2#3{Ann. Phys. (N.Y.) {\bf#1}, #2 (#3)}
\def \app#1#2#3{Acta Phys. Polonica {\bf#1}, #2 (#3)}
\def \arnps#1#2#3{Ann. Rev. Nucl. Part. Sci. {\bf#1}, #2 (#3)}
\def \cmts#1#2#3{Comments on Nucl. Part. Phys. {\bf#1}, #2 (#3)}
\def \cn{Collaboration}
\def \cp89{{\it CP Violation,} edited by C. Jarlskog (World Scientific,
Singapore, 1989)}
\def \efi{Enrico Fermi Institute Report No. EFI}
\def \f79{{\it Proceedings of the 1979 International Symposium on Lepton and
Photon Interactions at High Energies,} Fermilab, August 23-29, 1979, ed. by
T. B. W. Kirk and H. D. I. Abarbanel (Fermi National Accelerator Laboratory,
Batavia, IL, 1979}
\def \hb87{{\it Proceeding of the 1987 International Symposium on Lepton and
Photon Interactions at High Energies,} Hamburg, 1987, ed. by W. Bartel
and R. R\"uckl (Nucl. Phys. B, Proc. Suppl., vol. 3) (North-Holland,
Amsterdam, 1988)}
\def \ib{{\it ibid.}~}
\def \ibj#1#2#3{~{\bf#1}, #2 (#3)}
\def \ichep72{{\it Proceedings of the XVI International Conference on High
Energy Physics}, Chicago and Batavia, Illinois, Sept. 6 -- 13, 1972,
edited by J. D. Jackson, A. Roberts, and R. Donaldson (Fermilab, Batavia,
IL, 1972)}
\def \ijmpa#1#2#3{Int. J. Mod. Phys. A {\bf#1}, #2 (#3)}
\def \ite{{\it et al.}}
\def \jpb#1#2#3{J.~Phys.~B~{\bf#1}, #2 (#3)}
\def \lkl87{{\it Selected Topics in Electroweak Interactions} (Proceedings of
the Second Lake Louise Institute on New Frontiers in Particle Physics, 15 --
21 February, 1987), edited by J. M. Cameron \ite~(World Scientific, Singapore,
1987)}
\def \kdvs#1#2#3{{Kong.~Danske Vid.~Selsk., Matt-fys.~Medd.} {\bf #1}, No.~#2
(#3)}
\def \ky85{{\it Proceedings of the International Symposium on Lepton and
Photon Interactions at High Energy,} Kyoto, Aug.~19-24, 1985, edited by M.
Konuma and K. Takahashi (Kyoto Univ., Kyoto, 1985)}
\def \mpla#1#2#3{Mod. Phys. Lett. A {\bf#1}, #2 (#3)}
\def \nc#1#2#3{Nuovo Cim. {\bf#1}, #2 (#3)}
\def \np#1#2#3{Nucl. Phys. {\bf#1}, #2 (#3)}
\def \PDG{Particle Data Group, L. Montanet \ite, \prd{50}{1174}{1994}}
\def \pisma#1#2#3#4{Pis'ma Zh. Eksp. Teor. Fiz. {\bf#1}, #2 (#3) [JETP Lett.
{\bf#1}, #4 (#3)]}
\def \pl#1#2#3{Phys. Lett. {\bf#1}, #2 (#3)}
\def \pla#1#2#3{Phys. Lett. A {\bf#1}, #2 (#3)}
\def \plb#1#2#3{Phys. Lett. B {\bf#1}, #2 (#3)}
\def \pr#1#2#3{Phys. Rev. {\bf#1}, #2 (#3)}
\def \prc#1#2#3{Phys. Rev. C {\bf#1}, #2 (#3)}
\def \prd#1#2#3{Phys. Rev. D {\bf#1}, #2 (#3)}
\def \prl#1#2#3{Phys. Rev. Lett. {\bf#1}, #2 (#3)}
\def \prp#1#2#3{Phys. Rep. {\bf#1}, #2 (#3)}
\def \ptp#1#2#3{Prog. Theor. Phys. {\bf#1}, #2 (#3)}
\def \rmp#1#2#3{Rev. Mod. Phys. {\bf#1}, #2 (#3)}
\def \rp#1{~~~~~\ldots\ldots{\rm rp~}{#1}~~~~~}
\def \si90{25th International Conference on High Energy Physics, Singapore,
Aug. 2-8, 1990}
\def \slc87{{\it Proceedings of the Salt Lake City Meeting} (Division of
Particles and Fields, American Physical Society, Salt Lake City, Utah, 1987),
ed. by C. DeTar and J. S. Ball (World Scientific, Singapore, 1987)}
\def \slac89{{\it Proceedings of the XIVth International Symposium on
Lepton and Photon Interactions,} Stanford, California, 1989, edited by M.
Riordan (World Scientific, Singapore, 1990)}
\def \smass82{{\it Proceedings of the 1982 DPF Summer Study on Elementary
Particle Physics and Future Facilities}, Snowmass, Colorado, edited by R.
Donaldson, R. Gustafson, and F. Paige (World Scientific, Singapore, 1982)}
\def \smass90{{\it Research Directions for the Decade} (Proceedings of the
1990 Summer Study on High Energy Physics, June 25--July 13, Snowmass,
Colorado),
edited by E. L. Berger (World Scientific, Singapore, 1992)}
\def \tasi90{{\it Testing the Standard Model} (Proceedings of the 1990
Theoretical Advanced Study Institute in Elementary Particle Physics, Boulder,
Colorado, 3--27 June, 1990), edited by M. Cveti\v{c} and P. Langacker
(World Scientific, Singapore, 1991)}
\def \yaf#1#2#3#4{Yad. Fiz. {\bf#1}, #2 (#3) [Sov. J. Nucl. Phys. {\bf #1},
#4 (#3)]}
\def \zhetf#1#2#3#4#5#6{Zh. Eksp. Teor. Fiz. {\bf #1}, #2 (#3) [Sov. Phys. -
JETP {\bf #4}, #5 (#6)]}
\def \zpc#1#2#3{Zeit. Phys. C {\bf#1}, #2 (#3)}
\def \zpd#1#2#3{Zeit. Phys. D {\bf#1}, #2 (#3)}


\begin{thebibliography}{99}

\bibitem{Baym} G. Baym, {\it Lectures on Quantum Mechanics} (Benjamin, New
York, 1969).

\bibitem{GP} M. Gell-Mann and A. Pais, ``Behavior of neutral particles under
charge conjugation,'' \pr{97}{1387-9}{1955}.

\bibitem{CPT} J. Schwinger, ``The theory of quantized fields.  II,''
\pr{91}{713-28}{1953} (see esp.~p.~720 ff); ``The theory of quantized fields.
VI,'' \pr{94}{1362-84}{1954} (see esp.~Eq (54) on p.~1366 and p.~1376 ff); G.
L\"uders, ``On the equivalence of invariance under time reversal and under
particle-antiparticle conjugation for relativistic field theories,''
\kdvs{28}{5, 1-17}{1954}; ``Proof of the TCP theorem,'' \apny{2}{1-15}{1957};
W. Pauli, ``Exclusion principle, Lorentz group, and reflection of space-time
and charge,'' in W. Pauli, ed. {\it Niels Bohr and the Development of Physics}
(Pergamon, New York, 1955), pp.~30-51.

\bibitem{VA} R. P. Feynman and M. Gell-Mann, ``Theory of the Fermi
interaction,'' \pr{109}{193-8}{1958}; E. C. G. Sudarshan and R. E. Marshak,
``Chirality invariance and the universal Fermi interaction,''
\pr{109}{1860-2}{1958}.

\bibitem{CCFT} J. H. Christenson, J. W. Cronin, V. L. Fitch, and R. Turlay,
``Evidence for the $2 \pi$ decay of the $K_2^0$ meson,''
\prl{13}{138-40}{1964}.

\bibitem{GN} M. Gell-Mann, ``Isotopic spin and new unstable particles,''
\pr{92}{833-4}{1953}; ``On the Classification of Particles'' (1953,
unpublished); M. Gell-Mann, ``The interpretation of the new particles as
displaced charge multiplets,'' \nc{4}{Suppl.~848-66}{1956}; T. Nakano and K.
Nishijima, ``Charge independence for $V$-particles,'' \ptp{10}{581-2}{1953}; K.
Nishijima, ``Some remarks on the even-odd rule,'' \ptp{12}{107-8}{1954};
``Charge independence theory of $V$ particles,'' \ptp{13}{285-304}{1955}.

\bibitem{KL} K. Lande, E. T. Booth, J. Impeduglia, and L. M. Lederman,
``Observation of long-lived $V$ particles,'' \pr{103}{1901-4}{1956}.

\bibitem{KCP} T. D. Lee, R. Oehme and C. N. Yang, ``Remarks on possible
noninvariance under time reversal and charge conjugation,''
\pr{106}{340-5}{1957}; B. L. Ioffe, L. B. Okun' and A. P. Rudik, ``The problem
of parity non-conservation in weak interactions,''
\zhetf{32}{396-7}{1957}{5}{328-30}{1957}.

\bibitem{BW} B. Winstein, ``CP violation,'' in {\it Festi-Val -- Festschrift
for Val Telegdi}, ed.~by K. Winter (Elsevier, Amsterdam, 1988), pp.~245-265.

\bibitem{CL} T. P. Cheng and L. F. Li, {\it Gauge Theory of Elementary
Particles} (Oxford University Press, 1984).

\bibitem{Revs} See, e.g., K. Kleinknecht, ``CP violation in the $K^0 -
\bar K^0$ system,'' in \cp89, pp.~41-104, and references therein.

\bibitem{Sachs} See, e.g., R. G. Sachs, {\it The Physics of Time Reversal
Invariance} (University of Chicago Press, Chicago, 1988).

\bibitem{Kabir} See, e.g., P. K. Kabir, {\it The CP Puzzle} (Academic Press,
New York, 1968).

\bibitem{PDG} Particle Data Group, L. Montanet \ite, ``Review of particle
properties,'' \prd{50}{1173-1826}{1994}.

\bibitem{E731} Fermilab E731 \cn, L. K. Gibbons \ite, ``Measurement of the
CP violation parameter Re$(\epsilon'/\epsilon)$,'' \prl{70}{1203-6}{1993}.

\bibitem{NA31} CERN NA31 \cn, G. D. Barr \ite, ``A new measurement of direct
CP violation in the neutral kaon system,'' \plb{317}{233-42}{1993}.

\bibitem{Davies} K. Davies, {\it Ionospheric Radio} (Peter Peregrinus Ltd.,
London, 1990), p.~378.

\end{thebibliography}
\end{document}